\documentclass[a4paper, amsfonts, amssymb, amsmath, preprint, showkeys, nofootinbib]{revtex4}
\def\ps{\tilde{\psi}}
\def\D{\Delta}
\usepackage[english]{babel}
\usepackage[utf8]{inputenc}
\usepackage[colorinlistoftodos, color=green!40, prependcaption]{todonotes}
\usepackage{amsthm}
\usepackage{mathtools}
\usepackage{physics}
\usepackage{xcolor}
\usepackage{graphicx}
\usepackage[left=23mm,right=13mm,top=35mm,columnsep=15pt]{geometry} 
\usepackage{adjustbox}
\usepackage{placeins}
\usepackage[T1]{fontenc}
\usepackage{lipsum}
\usepackage{csquotes}
\begin{document}
\title{A case of thermodynamic failure in the Ginzburg--Landau approach to fluctuation superconductivity}

\author{Jorge Berger}
    \email[Correspondence email address: ]{jorge.berger@braude.ac.il}
    \affiliation{Department of Physics and Optical Engineering, Braude College, Karmiel, Israel}


\begin{abstract}
The Ginzburg--Landau approach postulates an energy density, together with an interpretation for the supercurrent, and invokes Ohm's law. We consider quasi-one-dimensional nonuniform superconducting loops, either smooth or piecewise uniform, that enclose a magnetic flux, above the critical temperature.
We evaluate the averages of the current and of the power per unit length delivered by the electric field due to thermal fluctuations. We consider three averages: canonical ensemble average, canonical ensemble in the reciprocal space, and time-average using a time-dependent model. All the evaluations imply that heat is imparted to part of the loop and removed in other part, despite the assumption that the loop is at uniform temperature.
\end{abstract}


\maketitle

\section{Introduction}  
The second law of thermodynamics asserts that within a system at uniform temperature there is no heat flux on average, no matter what its Hamiltonian is.

Here we examine the case of a non-uniform 1D superconducting loop that encloses a non-integer number of magnetic flux quanta, slightly above its critical temperature $T_c$ (a situation beyond the realms considered in Refs. \cite{Hirsch} or \cite{Nik}). The present study was motivated by Ref.\ \cite{2007}.

The questions which we intend to answer will be spelled out in the following two Sections; in Sections \ref{diag}--\ref{TDGL} we develop and apply three different techniques for the evaluation of the quantities of interest, and in Section \ref{last} we summarize our results.

\section{Formulation of the problem}
\subsection{Statistical average \label{IIA}}
If $q_1,...q_f$ are the coordinates of the phase space that describe the microstate of a system, $F(q_1,...q_f)$ is the energy of this system, and every part of the system is in equilibrium with a heat bath at temperature $T$, then the average value of any quantity $Q(q_1,...q_f)$ is predicted to be
\begin{equation}
    \langle Q\rangle =\left.\int Q(q_1,...q_f)e^{-F/k_BT}dq_1,...dq_f\middle/ \int e^{-F/k_BT}dq_1,...dq_f\right. \,,
    \label{Stat}
\end{equation}
where $k_B$ is Boltzmann’s constant and the integrals extend over the entire phase space.

\subsection{Energy of the system}
The system we consider consists of the electrons that are able to flow around a superconducting loop. For the sake of definiteness, let us also describe its environment: we may regard it as composed of four parts, i.e. the lattice of ions that form the loop, the fluid in which the loop is immersed, and the black body electromagnetic fields within the loop and in the fluid.

For a situation in which the temperature and the applied magnetic field are controlled, the static Ginzburg--Landau (GL) formalism invokes a complex field $\psi$, known as the ``order parameter,'' and the magnetic potential $\mathbf{A}$, which describe the microstate of a superconductor and play the role of the coordinates $q_1,...q_f$ in Sec.\ \ref{IIA}. GL attributes to superconductivity the free energy density \cite{Tinkham} $\alpha |\psi |^2+\beta |\psi |^4/2+(\hbar^2/2m)|(i\mathbf{\nabla}-2\pi \mathbf{A}/\Phi_0)\psi|^2$, where $\alpha$ and $\beta$ depend on the superconducting material, $m$ is the mass of a Cooper pair, $\Phi_0=\pi\hbar c/e$ is the quantum of magnetic flux and $e$ is the absolute value of the electron charge. $\alpha$ is an increasing function of temperature and vanishes for $T=T_c$. 

The total free energy of the system is the sum of the integral of the superconducting free energy density over the superconducting sample, the integral of the magnetic energy density over the entire space, and a term that does not depend on $\psi$ or $\mathbf{A}$. Ignoring fluctuations, the equilibrium values of $\psi$ and $\mathbf{A}$ are those that minimize the total free energy and are compatible with the constraints of the experimental setup \cite{Kopnin,Nolting}. Since the entropy is not assumed to depend on $\psi$ or $\mathbf{A}$, if the temperature is kept fixed, the difference between the free energy and the energy becomes an irrelevant additive constant. 

The GL free energy can be obtained from microscopic theory as an expansion, assuming that the order parameter is small. GL is intended to be a minimal model that retains only those terms that are essential. If $\alpha <0$, then the quartic term $\beta |\psi |^4/2$ has to be kept in order to obtain a free energy that is bounded from below. However, we will deal with a case in which $\alpha >0$ and therefore the quartic term will be dropped, except for Sec. \ref{quartic}.

We consider a superconducting loop of perimeter $L$, in which the linear dimensions of the cross section are much shorter than $L$ and than the typical distances over which $\psi$ and $\mathbf{A}$ vary; in this case the system becomes one dimensional and the position is described by the arc length $x$, $0\leq x\leq L$.

Although we consider a situation above the critical temperature, the order parameter does not vanish because of thermal fluctuations. Neglecting terms of order $O(|\psi |^4)$, the GL energy of the system can be written as 
\begin{equation}
  F=\int_0^L\left[\alpha |\psi |^2+
  \frac{\hbar^2}{2m}\left|\left(i\frac{d}{dx}-\frac{2\pi A}{\Phi_0}\right)\psi\right|^2\right]w(x)dx \,,
  \label{GLF}
\end{equation}
where \textit{A} is the tangential component of \textbf{A} and $w(x)$ is the cross section of the superconducting wire. 

We will assume that the magnetic self-inductance of the loop is negligible, as could be the case of a wide thin ribbon that surrounds a region with flattened cross section. As a consequence, the total magnetic flux will be just the applied flux. Consistently, we will assume that the contribution of the current (and therefore of $\psi$ and $A$) to the magnetic Gibbs free energy is negligible in comparison to $F$. 

The magnetic potential can be eliminated from (\ref{GLF}) by defining
\begin{equation}
    \ps (x)=\exp\left[\frac{2\pi i}{\Phi_0}\int_{x_1}^x A(x')dx'\right]\psi (x) \,,
    \label{tilde}
\end{equation}
where $x_1$ is arbitrary and $x$ can be extended beyond the range $0\leq x\leq L$ by identifying $x+L$ with $x$. Using (\ref{tilde}), (\ref{GLF}) becomes
\begin{equation}
    F=\oint [\alpha |\ps |^2+(\hbar^2/2m)|d\ps /dx|^2]w(x)dx \,.
    \label{GLps}
\end{equation}
It follows from (\ref{tilde}) that $\ps (x+L)=\ps (x)\exp (2\pi i \Phi /\Phi_0)$, with $\Phi =\oint A(x)dx$, so that $\ps$ is multivalued unless $\Phi /\Phi_0$ is integer.

Although the microstates of the system are described by two fields, $\psi$ and $A$, the accepted averaging procedure \cite{dt,Schmid,Sca1,Sca2,Var} includes only $\psi$ in the phase space.

Let us now fix some of the parameters in the system to be considered. The length $L$ of the loop, its electrical conductivity $\sigma$, its resistance $R$ (regarding the loop as an open circuit) and the temperature $T$ will be fixed. Unless explicitly stated, we will also fix $\alpha =\hbar^2/2mL^2$, so that the coherence length (the length over which $\psi$ changes significantly) will equal the perimeter of the loop. 
As mentioned above, the total flux will be the applied flux, which we will fix as 
$\Phi =\Phi_0/4$, so that $\ps (x+L)=i\ps (x)$. These values define a typical cross section $w_0$ and a dimensionless quantity $\gamma$:
\begin{equation}
    w_0=L/\sigma R\,,\;\;\gamma =\hbar /e^2R\,.
    \label{mensionless}
\end{equation}

Several possibilities for the cross section $w(x)$ will be considered. A smooth profile will be 
\begin{equation}
    w(x)=(2w_0/\sqrt{3} )[1-\cos (2\pi x/L)/2] \,;
    \label{smooth}
\end{equation} 
a family of discrete profiles will be described in the following section.

\subsection{Discrete profiles}
We divide the loop into $N$ segments of length $L/N$, centered at $x=x_k:=(\lambda +k-1)L/N$ with $0\leq \lambda < 1$ and $k=1,...N$. The cross section of segment $k$ will be
\begin{equation}
    w_k=(w_0/N)[1-\cos (2\pi x_k/L)/2 ]\sum_{j=1}^N[1-\cos (2\pi x_j/L)/2]^{-1}\,.
\end{equation}
In the limit of large $N$, this profile becomes the smooth profile (\ref{smooth}).

Let us now build a model for the energy of this system, motivated by the GL energy (\ref{GLps}). Instead of the field $\ps (x)$, we assign the value $\ps_k$ to the entire segment $k$; the derivative $d\ps /dx$ is replaced by the finite difference $N(\ps_{k+1}-\ps_k)/L$ and we define
\begin{equation}
  F_{N,\lambda}:=\sum_{k=1}^N\left\{ \frac{\alpha Lw_k}{N}|\ps_k |^2+\frac{\hbar^2Nw_{k_+}}{2mL}|\ps_{k+1}-\ps_k |^2  \right\} \,,
  \label{Fnlam}
\end{equation}
where $w_{k_+}$ is some average between $w_k$ and $w_{k+1}$, and $\ps_{N+1}=\exp (2\pi i\Phi /\Phi_0)\ps_1$. We have studied two possibilities for $w_{k_+}$: $w_{k_+}=(w_k+w_{k+1})/2$ and $w_{k_+}=2(w_k^{-1}+w_{k+1}^{-1})^{-1}$.

We still require discretizations for $\psi$ and $A$. We assign the magnetic potential $A_k^+$ (respectively $A_k^-$) to the half-segment $x_k<x<x_k+L/2N$ (respectively $x_k>x>x_k-L/2N$)  and define
\begin{equation}
    \ps_1:=\psi_1 \,;\;\; \ps_k:=\exp\left[\frac{\pi iL}{N\Phi_0}\left(A_1^++A_k^-+\sum_{j=2}^{k-1}(A_j^-+A_j^+)\right)\right]\psi_k \;\; 2\leq k\leq N\,.
    \label{pspsi}
\end{equation}
In terms of $\psi$ and $A$, the energy is
\begin{equation}
      F_{N,\lambda}=\sum_{k=1}^N\left\{ \frac{\alpha Lw_k}{N}|\psi_k|^2+\frac{\hbar^2Nw_{k_+}}{2mL}\left(|\psi_k |^2+|\psi_{k+1} |^2-2\mathrm{Re}\left[\psi_{k+1}\psi_k^*e^{(\pi iL/N\Phi_0)(A_k^++A_{k+1}^-)}\right]\right)  \right\}\,.
      \label{Fpsi}
\end{equation}

Although the systems described here approach the smooth loop only in the limit of large $N$, any of them, for any $N$, $\lambda$ and $\{ w_{k_+} \}$, can be regarded as a self-consistent model that should obey the laws of thermodynamics.

\section{Quantities of Interest}
Unlike ``classical'' equilibrium-thermodynamics, statistical mechanics can deal not only with state functions, but also with transport quantities such as the total current and heat transport. Expressions for these quantities will be obtained in this section.

In the case of the smooth loop, the supercurrent density is given by \cite{Tinkham}
\begin{equation}
    J_S(x)=-\frac{2e\hbar}{m}\mathrm{Im}\left[ \psi^*\left(\frac{d}{dx}+\frac{2\pi i}{\Phi_0}A\right)\psi\right] =-\frac{2e\hbar}{m}\mathrm{Im}\left[ \ps^*\frac{d\ps}{dx}\right] \,.
    \label{Js}
\end{equation}

Although the expression for $J_S$ looks the same as the expected current density of a charged particle described by the Schroedinger equation, we should note that in the Schroedinger case this expression follows from charge conservation, whereas in the GL case it follows from variation of the energy functional with respect to $A(x)$. Similarly, in the discrete case, in the half-segment where the magnetic potential is $A_k^\pm$, the supercurrent density is given by $J_{Sk}^\pm =-(2Nc/Lw_k)\partial F_{N,\lambda}/\partial A_k^\pm$, where the meaning of partial derivative is that the magnetic potential in the other regions of the loop, and also $\{ \psi_j\}$, are kept constant. Using (\ref{Fpsi}) and (\ref{pspsi}) we obtain
\begin{equation}
J_{Sk}^+=- (2Ne\hbar w_{k_+} /mLw_k){\rm Im}[\ps_k^*\ps_{k+1}]\;;\;\;J_{Sk}^-=- (2Ne\hbar w_{(k-1)_+} /mLw_k){\rm Im}[\ps_{k-1}^*\ps_k] \;.
\label{JSKn}
\end{equation}

We denote by $I$ the total current around the loop. As usual, we assume electroneutrality, so that $I$ does not depend on $x$. By Ohm's law, the electric field is $E(x)=[I/w(x)-J_S(x)]/\sigma$ in the smooth case and
\begin{equation}
    E_k^\pm =(I/w_k-J_{Sk}^\pm )/\sigma
    \label{field}
\end{equation}
in the discrete case. Since the magnetic flux through the loop remains constant, the circulation of the electric field vanishes and the current is
\begin{equation}
I=\left. \oint J_Sdx\middle/ \oint w^{-1}dx\right.
\label{Ismooth}
\end{equation}
 in the smooth case and, noting that $\sum_{k=1}^N w_k^{-1}=N/w_0$, 
\begin{equation}
    I=\frac{w_0}{2N}\sum_{k=1}^N(J_{Sk}^-+J_{Sk}^+) 
    \label{current}
\end{equation}
in the discrete case.

The power per unit length delivered by the electric field to the flowing charges is
\begin{equation}
    P(x)=IE(x) \,.
    \label{powlength}
\end{equation}
If this power is not taken somewhere else by the supercurrent, it has to be passed as heat 
to the environment. We note that, since $I$ is uniform and $E(x)$ is conservative, the total power delivered by the electric field is zero.

For the sake of definiteness, let us focus our attention on a segment of the lattice of ions close to the position $x$. If a fraction $\kappa_1$ of the work performed by the electric field originates from this segment, and a fraction $\kappa_2$ of the heat released by the flowing charges is passed to this segment, then the internal energy of the segment per unit length and time will increase by $(\kappa_2-\kappa_1)P$. This energy cannot accumulate at the segment and, since the environment is at uniform temperature, implying that there will be no heat flow on the average, the average of $(\kappa_2-\kappa_1)P$ has to vanish. $\kappa_1$ and $\kappa_2$ are not expected to equal one another, because they involve different mechanisms, e.g. the electric field at the segment is influenced by distant segments \cite{Johnson}, and this field performs work also on Cooper pairs. Therefore, the thermodynamic requirement is that the average of $P(x)$ has to vanish at every position along the loop. 

Since the expression (\ref{GLps}) for the energy penalizes the gradient of $\ps$, at first glance one might think that this gradient vanishes on the average, leading to $\langle I \rangle =0$, but the gradient cannot vanish because generically $\ps (x+L)\neq\ps (x)$. If the energy density in (\ref{GLps}) were not weighed by $w(x)$, then, by symmetry, the average of this gradient would have uniform size. This uniform situation was considered in \cite{2007} and in this case  $\langle P \rangle =0$. In the present situation, the smaller $w(x)$, the larger the fluctuations and the resistance per length near $x$, and we should anticipate more complex behavior.

\section{Evaluation of averages for discrete systems}\label{diag}
The sets $\{ \psi_k\}$ and $\{ \ps_k\}$ cover the same phase space, so that it makes no difference over which of them we average. For further simplification, we diagonalize (\ref{Fnlam}) numerically, taking the form
\begin{equation}
    F_{N,\lambda}=\sum_{k=1}^N f_k |\varphi_k|^2 \,.
\end{equation}
The passage from $\{ \ps_k\}$ to $\{\varphi_k\}$ is a rotation in the phase space, and any of them can be used for averaging.

The electric field in (\ref{field}) and the current in (\ref{current}) are quadratic expressions. $\langle\varphi_j^*\varphi_i\rangle$ vanishes for $j\neq i$, and $\langle\varphi_k^*\varphi_k\rangle =k_BT/f_k$. We found $\langle E_k^\pm\rangle =0$ in all the cases considered and the values of the average currents obtained in this way are shown in the first two rows of Table \ref{tablecurr}. Note that the current can be expressed solely in terms of $ek_BT/\hbar$ because $\alpha$ has been fixed. 

\begin{table*}[tbhp]
\caption{\centering Average current in units of $ek_BT/\hbar$. The `method'-column directs to the section where the evaluation method is described. `a' stands for the arithmetic mean $w_{k_+}=(w_k+w_{k+1})/2$ and `h' for the harmonic mean  $w_{k_+}=2(w_k^{-1}+w_{k+1}^{-1})^{-1}$. The pairs of numbers in the headings row stand for $N$ and $\lambda$.} Unless explicitly mentioned, $\alpha =\hbar^2/2mL^2$ and $\beta =0$ in all figures and tables.
\begin{center}
\begin{tabular}{ccccccccc}
\hline
method&$w_{k_+}$&3,0&3,0.5&4,0&4,0.25&5,0&10,0&15,0\\
\hline
\ref{diag} &a&-1.291&-1.286&-1.271&-1.270&-1.259&-1.238&-1.234\\
\ref{diag}&h&-1.215&-1.252&-1.227&-1.232&-1.230&-1.230&-1.230\\
\ref{TDGL}&h&-1.214&-1.251&-1.224&-1.228&-1.229&-1.225&-1.228\\
\hline
\end{tabular}
\end{center}
\label{tablecurr}
\end{table*}

\begin{figure}
\scalebox{0.85}{\includegraphics{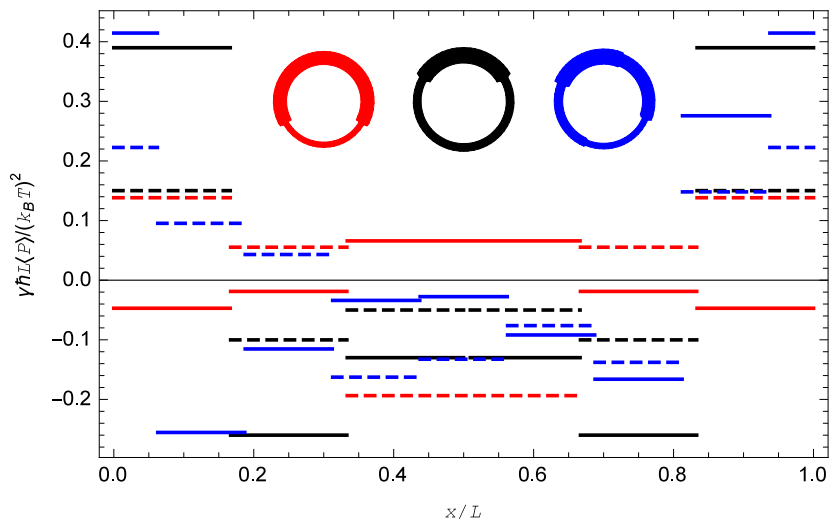}}
\caption{\label{smallN} Average power passed to the flowing charges per unit length [Eq.\ (\ref{powlength})], as a function of position, evaluated as described in Sec.\ \ref{diag}, for small values of $N$. The
icons at the top of the figure depict the cross section of the system considered, with the position where $x=0$ and $x=L$ join at the bottom of the icon. Solid lines stand for $w_{k_+}=(w_k+w_{k+1})/2$ and dashed lines for $w_{k_+}=2(w_k^{-1}+w_{k+1}^{-1})^{-1}$. Red: $N=3$, $\lambda =0$; black: $N=3$, $\lambda =0.5$; blue: $N=4$, $\lambda =0.25$. $\gamma$ is defined in Eq.\ (\ref{mensionless}).}
\end{figure}

The average power per unit length, $\langle P_k^\pm\rangle$, is obtained from Eq.\ (\ref{powlength}). $IE_k^\pm$ is a long linear combination of components of the form $\varphi_i^*\varphi_j\varphi_\ell^*\varphi_m$. Most of these components give no contribution to the average, and the only contributions are those of the forms $\langle (\varphi_i^*\varphi_i)^2\rangle =2(k_BT/f_i)^2$ and $\langle \varphi_i^*\varphi_i\varphi_j^*\varphi_j\rangle =(k_BT)^2/f_if_j$ for $i\neq j$. Representative results obtained for small values of $N$ are shown in Fig.\ \ref{smallN}. From the second law of thermodynamics and the fact that the temperature is uniform, we would expect no local energy flow from the system to the heat bath on average, i.e.\ $\langle P_k^\pm\rangle$ should vanish identically (as would be the case if $I$ and $E_k^\pm$ were uncorrelated). Instead, we see in Fig.\ \ref{smallN} that, with the exception of the case $(N,\lambda )=(3,0)$ with arithmetic mean, the flowing charges receive power from the electric field at the thinnest  part of the loop and lose it at the broadest part.

\begin{figure}
\scalebox{0.85}{\includegraphics{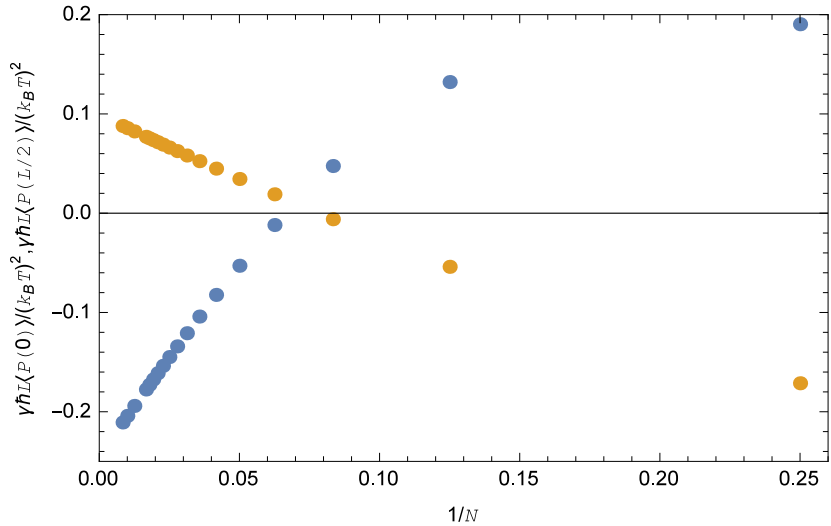}}
\caption{\label{P0L2} Average power per unit length delivered by the electric field to the flowing charges at the thinnest (blue) and at the broadest (orange) segment of the loop, as functions of $N$. In all cases $N$ is an integer multiple of 4, $w_{k_+}=2(w_k^{-1}+w_{k+1}^{-1})^{-1}$ and $\lambda =0$. When passing fron $N=12$ to $N=16$ the signs of $\langle P(0)\rangle$ and $\langle P(L/2)\rangle$ reverse.}
\end{figure}

We would like to know whether this disagreement with thermodynamics is a feature of discretization or a property of the GL approach. With this purpose, we investigate the behavior of $\langle P_k^\pm\rangle$ as $N$ increases and the smooth limit is approached. Judging by the $N$-dependence of $\langle I\rangle$, we expect a faster convergence when $\{ w_{k_+}\}$ are taken as harmonic means. (The values of $\langle I\rangle$ for $N=100$ and $N=120$ coincide within seven significant figures.) The results are shown in Fig.\ \ref{P0L2}. Surprisingly, the effect reverses at intermediate values of $N$ and, as the smooth limit is approached, the flowing charges receive power at the broadest part of the loop and lose it at the thinnest part.

The strong $N$-dependence of $\langle P(x)\rangle$, that can even reverse its sign, is related to the coherence length. If we diminish the coherence length by a factor of 2, by setting $\alpha =2\hbar^2/mL^2$, then the crossover found in Fig.\ \ref{P0L2} occurs between $N=4$ and $N=8$. In the opposite direction, if the coherence length increases by a factor of 2, then the crossover occurs between $N=36$ and $N=40$ for $\langle P(L/2)\rangle$, and between $N=48$ and $N=52$ for $\langle P(0)\rangle$. In both cases, the larger the coherence length, the larger the tendency for power capture by the field at the broad part of the loop and power delivery by the field at the thin part. The $N$-dependence of the sign of $\langle P\rangle$ can be therefore interpreted as an effective increase of the coherence length due to the requirement of a uniform value of $\ps (x)$ along each entire segment; the smaller $N$ is, the longer the segments and the larger the effective coherence length. The impact of the coherence length will be examined again, for the case of a smooth loop, in Sec.\ \ref{smoothF}.  

\begin{figure}
\scalebox{0.85}{\includegraphics{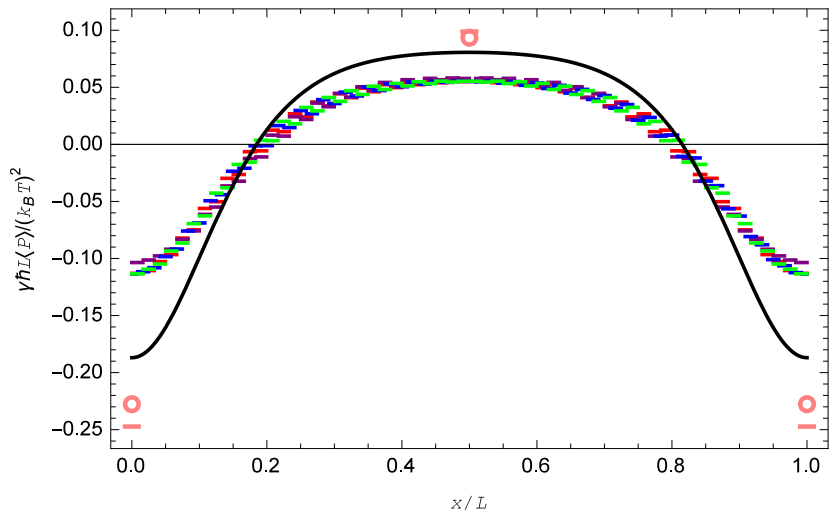}}
\caption{\label{d30}The colored segments show the average power per unit length as a function of position, obtained as described in Section \ref{diag}, for $N=30$. The purple lines were evaluated taking $w_{k_+}$ as an arithmetic mean and the other lines taking $w_{k_+}$ as a harmonic mean. Purple: $\lambda =0$; red: $\lambda =0$; blue: $\lambda =0.25$; green: $\lambda =0.5$. The pink segments at $x=0$, $x=L/2$ and $x=L$ mark the limit $N\to\infty$, obtained from extrapolation in Fig.\ \ref{P0L2}. The black curve and the pink circles were obtained as explained in Section \ref{smoothF}, with $M=15$ for the curve and $M=50$ for the circles.}
\end{figure}

Figure \ref{d30} enables us to appraise the influence of discretization on the power delivered, for a moderately large value of $N$. All the short segments that delineate a bell shape were obtained for $N=30$, but using different averages for $w_{k_+}$ or different orientations $\lambda$. The fact that all the results nearly coalesce indicates that discretization has no qualitative influence, and suggests that, if $\langle P(x)\rangle$ does not vanish for moderately large $N$, it will also not vanish for the smooth loop. 

\section{Averaging in the reciprocal space \label{smoothF}}
In the case of a smooth profile we can write $\ps$ as a modified Fourier series, 
\begin{equation}
   \ps (x)= \sum b_n e^{2\pi i(n+1/4)x/L}\;.
   \label{Four}
\end{equation}
This sum extends over all integers, but we will take $|n|\le M$, where $M$ will be increased until acceptable convergence is obtained.

From (\ref{Js}) and (\ref{Four}),
\begin{equation}
    J_S(x)=-\frac{4\pi e\hbar}{mL}\mathrm{Re}\left[\sum_{-M\le\ell ,n\le M} b_\ell^* b_n \left(n+\frac{1}{4}\right) e^{2\pi i(n-\ell )x/L}\right] \,;
    \label{Jss}
\end{equation}
from (\ref{Ismooth}) and (\ref{Jss}), using (\ref{smooth}),
\begin{equation}
    I=-\frac{4\pi e\hbar w_0}{mL}\sum_{-M\le n\le M}  \left(n+\frac{1}{4}\right)b_n^* b_n\,;
\end{equation}
and from (\ref{GLps}), (\ref{smooth}) and (\ref{Four})
\begin{eqnarray}
    F&=&\frac{2w_0L}{\sqrt{3}}\left\{ \sum_{n=-M}^M \left[\alpha +\frac{(2\pi\hbar )^2}{2mL^2}\left(n+\frac{1}{4}\right)^2\right]b_n^* b_n \right.   \nonumber \\
     &-&\left.\frac{1}{4} \sum_{n=-M}^{M-1}\left[\alpha +\frac{(2\pi\hbar )^2}{2mL^2}\left(n+\frac{1}{4}\right)\left(n+\frac{5}{4}\right)\right](b_{n+1}^* b_n+b_{n+1} b_n^*) \right\}\;.
     \label{Fbb}
\end{eqnarray}

We can now proceed as in Section \ref{diag}. We diagonalize (\ref{Fbb}), so that it takes the form $F=\sum_{n=-M}^M f_n|c_n|^2$. It follows that the only quadratic terms with nonzero average are $\langle c_j^*c_j\rangle =k_BT/f_j$, and the only quartic terms that contribute are those of the forms $\langle (c_i^*c_i)^2\rangle =2(k_BT/f_i)^2$ and $\langle c_i^*c_ic_j^*c_j\rangle =(k_BT)^2/f_if_j$ for $i\neq j$. Finally, the average of the power per length in (\ref{powlength}) is evaluated as
\begin{equation}
    \langle P(x)\rangle =\frac{\hbar w_0 \langle I[I-J_S(x)w(x)]\rangle}{\gamma e^2Lw(x)}\,.
    \label{PFour}
\end{equation}

The values of the average current obtained in this way converge slowly for increasing $M$. For $M=5$ (respectively 10, 15, 20) we obtain $\hbar\langle I\rangle /ek_BT=-1.289$ (respectively -1.261, -1.251, -1.246).

The values of $\langle P(x)\rangle$ obtained from (\ref{PFour}) for $M=15$ are shown by the black curve in Fig.\ \ref{d30}. This curve is not expected to coincide with the results obtained in Section \ref{diag}, because the limits $M,N\to\infty$ have not been reached. Nevertheless, this figure strongly suggests that all the results would coincide in this limit and rebuts the possibility that $\langle P(x)\rangle\equiv 0$.

\begin{figure}
\scalebox{0.85}{\includegraphics{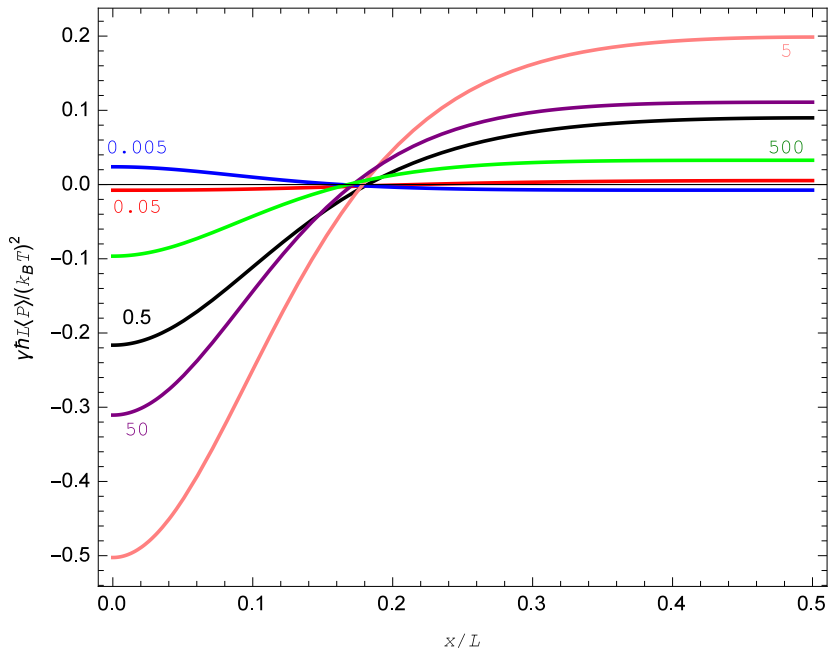}}
\caption{\label{coherence}Powers delivered per unit length, as functions of position, for a smooth loop. The values of $\alpha mL^2/\hbar^2$ are shown next to each line. Since $\langle P(L-x)\rangle =\langle P(x)\rangle$, only the range $0\le x\le L/2$ is shown.}
\end{figure}

The analysis of the smooth loop enables us to disentangle the influence of the coherence length from that of the segments' length found in Section \ref{diag}. Figure \ref{coherence} shows the delivered powers as functions of position for a wide range of values of $\alpha$. As expected, these powers vanish for coherence length much shorter than $L$ and saturate for coherence length much larger than $L$. As in Section \ref{diag}, the electric field delivers energy to the flowing charges in the thin (respectively broad) part of the loop when the coherence length is sufficiently large (respectively short), and the crossover occurs for $\alpha\sim 0.05\hbar^2/mL^2$. The largest energy transfers are found for a coherence length about a third of the perimeter of the loop.  

The evidence of this and the previous  section leads us to conclude that the GL approach does not comply with thermodynamics, neither for a discrete nor for a smooth nonuniform loop. We might think of several reservations to this conclusion: (i) Johnson noise has to be added to Eq.\ (\ref{field}); (ii) the power that the flowing charges release should not be interpreted as energy transferred to the heat bath, but rather as potential energy that is converted to kinetic energy of the moving charges and then transported as an energy supercurrent; (iii) the magnetic potential $A$ should be included in the phase space; (iv) besides the work that the electric field can perform on the charges, our system can also exchange energy by breaking Cooper pairs (i.e. by changing the value of $|\psi (x)|$); and (v) the currents shown in Table \ref{tablecurr} do not become negligible in the limit $w_0\rightarrow 0$. Moreover, $\psi$ has to diverge in the limit of small cross section and therefore the quartic term in the free energy density cannot be neglected. In order to address these reservations we require a model for time evolution. 

\section{Time dependent model \label{TDGL}}
In this section we discretize time into steps $\D t$, sufficiently small to justify regarding the changes of $\psi$ and $A$ during a step as infinitesimal.
\subsection{Johnson noise and evolution equations\label{JN}}
During a step $\Delta t$, the Johnson noise adds to the electric field in Eq.\ (\ref{field}) an additional term $\eta_{k,\pm}^A\sqrt{4k_BTN/\sigma w_kL\Delta t}$, where $\eta_{k,+}^A$ and $\eta_{k,-}^A$ stand for random variables with zero average, variance 1 and normal distribution. Choosing a gauge such that $cE_k^\pm=-\Delta A_k^\pm/\Delta t$, where $\Delta A_k^\pm(t)=A_k^\pm(t+\Delta t)-A_k^\pm(t)$ and using (\ref{mensionless}), Eq.\ (\ref{field}) becomes an evolution equation for $A_k^\pm(t)$:
\begin{equation}
    \Delta A_k^\pm=-(cw_0\hbar /\gamma e^2L)(I/w_k-J_{Sk}^\pm)\Delta t
    -(2c/eL)\eta_{k,\pm}^A\sqrt{k_BT\hbar Nw_0\Delta t /\gamma w_k}\,.
    \label{evA}
\end{equation}

Equation (\ref{evA}) is implied by the time-dependent Ginzburg--Landau model (TDGL), which has Ohm's law built into it. TDGL is the simplest dynamical model that converges to GL when equilibrium is attained. 
Ignoring fluctuations, for a system in which there are no supercurrents in equilibrium (provided that fluctuations are ignored), and taking a gauge with no electrical potential, TDGL states that $d\psi /dt\propto -\delta F/\delta\psi^*$ and $dA /dt\propto -\delta F/\delta A$ \cite{Schmid1,Kopnin}, or, in discretized form, $\D\psi_k /\D t\propto -w_k^{-1}\partial F_{N,\lambda}/\partial\psi_k^*$ and $\D A_k^\pm /\D t\propto -w_k^{-1}\partial F_{N,\lambda}/\partial A_k^\pm\propto J_{Sk}^\pm$. From the equation $\D A_k^\pm /\D t\propto J_{Sk}^\pm$, subjected to the constriction $\sum_{k=1}^N(A_k^-+A_k^+)=$ constant, and adjusting the constant of proportionality to the conductivity of the loop, we recover Eq.\ (\ref{evA}) without the stochastic term. The last term is a Langevin term, the variance of which is determined by the fluctuation-dissipation theorem \cite{Lang}.

Similarly, from the equation $\D\psi_k /\D t\propto -w_k^{-1}\partial F_{N,\lambda}/\partial\psi_k^*$,
\begin{eqnarray}
    \frac{\D\psi_k}{\D t}=&-&\frac{\Gamma\alpha}{\hbar}\psi_k-\frac{N^2\hbar\Gamma}{2mL^2w_k}[w_{k_+}(\psi_k-\psi_{k+1}e^{(\pi iL/N\Phi_0)(A_k^++A_{k+1}^-)})\nonumber \\
    &+&w_{(k-1)_+}(\psi_k-\psi_{k-1}e^{-(\pi iL/N\Phi_0)(A_{k-1}^++A_k^-)})]+\eta_k^\psi\sqrt{\frac{N\Gamma k_BT}{\hbar Lw_k\D t}}\;,
    \label{Dpsi}
\end{eqnarray}
where $\Gamma$ is a dimensionless material constant and both the real and the imaginary part of $\eta_k^\psi$ are random variables with zero average, variance 1, and normal distribution. Using (\ref{pspsi}) and assuming that $\D t$, $\D A_j^\pm$ and $\D\psi_k$ are sufficiently small, (\ref{Dpsi}) gives the evolution of $\ps_k$:
\begin{eqnarray}
 \ps_k(t+\Delta t)&=&\left\{ \left(1-\frac{\Gamma\alpha\Delta t}{\hbar}\right)\ps_k(t)+\frac{N^2\hbar\Gamma\Delta t}{2mL^2w_k}[w_{(k-1)_+}\ps_{k-1}(t)\right. \nonumber \\
 &-&\left. (w_{(k-1)_+}+w_{k_+})\ps_k(t)+w_{k_+}\ps_{k+1}(t)]+\eta_k^\psi\sqrt{\frac{N\Gamma k_BT\Delta t}{\hbar Lw_k}}\right\}{\rm prod}_k\,,
 \label{evpsi}
\end{eqnarray}
with
\begin{equation}
    {\rm prod}_1=1\;;\;\;{\rm prod}_{j+1}={\rm prod}_j\exp [iLe(\D A_j^++\D A_{j+1}^-)/Nc\hbar]\,.
\end{equation}

TDGL is valid for gapless superconductivity. This is the situation that we are considering, since $\alpha >0$. At any rate, in this study we are concerned only with the question of compatibility of TDGL with thermodynamics, and not with other possible limitations. We point out that, since we are considering uniform temperature, we do not have to deal with heat diffusion, as in Eq.\ (4) of \cite{Boris}.

We note that $\gamma$ and $\D t$ enter (\ref{evA}) only through their ratio; likewise, $\Gamma$ and $\D t$ enter (\ref{evpsi}) only through their product. Therefore, changing $\gamma$ and $\Gamma$ while keeping their product unchanged leads to the same evolution, although at a different rate. Anyway, in this study we are interested only in equilibrium values, which should not be affected by the choices of $\gamma$ and $\Gamma$.

Unlike the procedure in Sec.\ \ref{diag}, which did not include the magnetic potential in the phase space, in TDGL the fields $A$ and $\psi$ stand on the same footing.

\subsection{Energy current and pair creation energy\label{Ee}}
We may decompose $F_{N,\lambda}$ in (\ref{Fnlam}) into a sum of terms $F_k$ and $F_{k_+}$, where $F_k$ is the term that contains $|\ps_k|^2$ and $F_{k_+}$ contains $|\ps_{k+1}-\ps_k|^2$. We may regard $F_k$ as located in segment $k$ and $F_{k_+}$ as located at the boundary between $k$ and $k+1$.

A change $\D\ps_k$ in $\ps_k$ leads to changes $\D F_k$, $\D F_{k_+}$ and $\D F_{(k-1)_+}$ in $F_k$, $F_{k_+}$ and $F_{(k-1)_+}$. The time average of ${\rm Re}[\ps_k^*\D \ps_k]$ vanishes, and accordingly we take only cross terms into account. $\D F_{k_+}+\D F_{(k-1)_+}$ may be interpreted as the change in the energy in segment $k$ due to pair production, whereas $I_{Ek}\D t=(\D F_{k_+}-\D F_{(k-1)_+})/2$ may be interpreted as the energy transported by the supercurrent from the negative to the positive boundary of segment $k$. From (\ref{Fnlam}) we obtain
\begin{equation}
    I_{Ek}\D t=(\hbar^2N/2mL){\rm Re}[\D\ps_k(w_{(k-1)_+}\ps_{k-1}-w_{k_+}\ps_{k+1})^*] \,.
    \label{IE}
\end{equation}
Equation (\ref{IE}) can be identified as a discretized version of Eq. (D) in Ref. \cite{Schmid1}, adapted to the case of a nonuniform wire.

\subsection{Results for Sections \ref{JN} and \ref{Ee}}
The averages in this Section are not obtained from Eq.\ (\ref{Stat}), but are rather time-averages. We took $\gamma \Gamma =1$ and, for evaluation of the average current, we took $\D t =2\times 10^{-4}\gamma\hbar/k_BT$ and followed the evolution equations (\ref{evA}) and (\ref{evpsi}) during $2\times 10^{10}$ steps for $N<10$\, ($10^{10}$ steps for $N\ge 10$). The initial value of $\ps$ was zero, and the $10^7$ initial steps were not included in the average. Our results for the current are shown in the last row in Table \ref{tablecurr}.

Comparison of the last two rows in Table \ref{tablecurr} shows that these two apparently independent procedures lead to the same results, but the time-averages have a random uncertainty, slighly larger than $10^{-3}ek_BT/\hbar$, that can be attributed to the fact that the averaging spanned time was not infinite. There is also a systematic discrepancy of a similar size, that can be attributed to the fact that $\D t$ was not infinitesimal.

Evaluation of $\langle I_E\rangle$ and $\langle P\rangle$ required longer periods of time and was limited to small values of $N$. For $N=5$, the evolution of $\ps$ and $A$ was followed during $1.6\times 10^{11}$ steps of $\D t=2\times 10^{-3}\gamma\hbar/k_BT$ and, for $N=8$, during $10^{11}$ steps of $\D t =10^{-3}\gamma\hbar/k_BT$. For every $k$,
$\langle \D F_{k_+}\rangle /\D t$ vanished within its statistical uncertainty, which was of the order of $10^{-4}(k_BT)^2/\gamma\hbar$ and therefore the energy supercurrent and the pair production energy had no significant influence on the transferred powers.

\begin{figure}
\scalebox{0.85}{\includegraphics{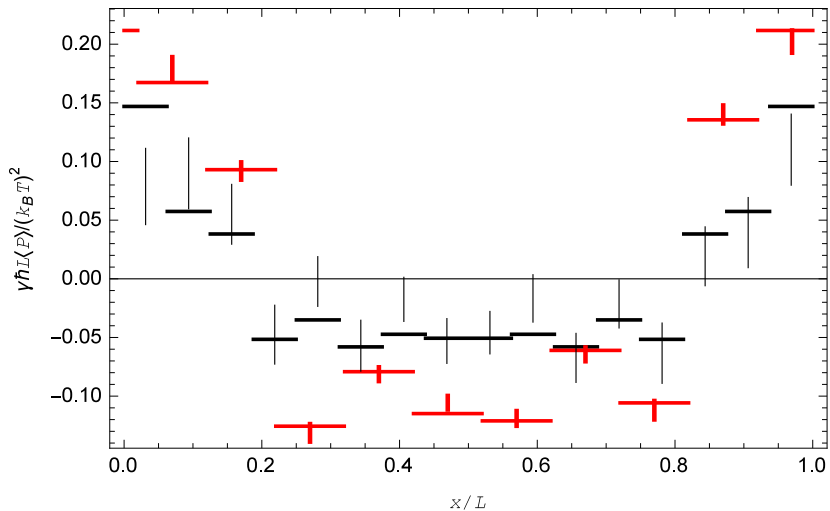}}
\caption{\label{comparison}Powers per unit length delivered by the electric field to the flowing charges, as functions of position. The horizontal lines were evaluated using Eq.\ (\ref{Stat}) and the vertical lines are time-averages, obtained as explained in Section \ref{TDGL}. The vertical lines are centered at the time-averages and their half-lengths equal the standard deviation divided by the square root of the number of steps. The horizontal lines are free of statistical uncertainty. Black: $N=8$, $\lambda =0.5$; red: $N=5$, $\lambda =0.1$. $w_{k_+}$ was taken as the harmonic mean of $w_{k}$ and $w_{k+1}$.}
\end{figure}

Figure \ref{comparison} compares the local delivered powers found in this Section with those obtained using Eq.\ (\ref{Stat}). Within the expected statistical uncertainty, the results obtained by both procedures agree with each other in all the cases, and in most cases they are convincingly different from zero.

Figure \ref{comparison} shows that Johnson noise does not clear away the local power transfer. Moreover, if we use (\ref{field}) instead of (\ref{evA}), we obtain different results for $\langle I\rangle$, $\langle I_E\rangle$ and $\langle P\rangle$, indicating that this noise can actually be regarded as a source of the power transfer distribution.

\subsection{Influence of the quartic term \label{quartic}}
When fluctuations are present, any coherent part of the system acquires energies of the order of $k_BT$. The smaller $w_0$, the smaller the volume of these parts and the larger the energy densities and the values that $|\psi |$ has to reach. However, GL relies on the premise that the energy density can be expanded in powers of $|\psi |$. Therefore, the situation we are considering is such that $w_0$ is sufficiently small to permit a 1D treatment, but sufficiently large to permit invoking GL.

In this Section we want to examine whether the nonzero powers that we have found are an artifact of the quadratic form of the energy density that we have kept, and will be wiped out if the quartic term is not neglected. Including this term is straightforward: all we have to do is replace $\alpha$ with $\alpha +\beta |\ps |^2$ in Eqs.\ (\ref{Dpsi}) and (\ref{evpsi}).

Table \ref{tquartic} displays the results that we have obtained for several values of $\alpha$ and $\beta$, for a particular discrete loop and for the particular temperature $T=\hbar^2/k_BmL^2$. We present only one ``power per length,'' that was evaluated in the four half-segments in the region $(\lambda -1)L/N\le x\le (\lambda +1)L/N$, at the thin part of the loop.

\begin{table*}[tbhp]
\caption{\centering Current and power per length for the profile $N=5$, $\lambda =0.1$, for temperature $T_0=\hbar^2/k_BmL^2$. $\alpha$ is in units of $\hbar^2/mL^2$, $\beta$ is in units of $\hbar^2 w_0/mL$ and the power per length is in units of $(k_BT_0)^2/\gamma\hbar L$. The power per length was evaluated in the region $-0.18\le x/L\le 0.22$ and has a statistical uncertainty $\sim 0.004$. The global correlation is defined as $\prod_{j=1}^N [|\langle\ps_j^*\exp(-2\pi i\Phi /N\Phi_0)\ps_{j+1}\rangle |/\langle |\ps_j\ps_{j+1}|\rangle]$}.
\begin{center}
\begin{tabular}{|c|c|c|c|c|}
\hline
$\alpha$&$\beta$&$\langle I\rangle$&power per length&global correlation\\
\hline
-0.5 &1&-1.076&0.133&0.392\\
0 &1&-0.830&0.095&0.331\\
0 &2&-0.571&0.057&0.251\\
0.005&0&-1.971&0.297&0.548\\
0.5&0&-1.222&0.153&0.426\\
0.5&0.5&-0.827&0.087&0.331\\
0.5&1&-0.652&0.061&0.279\\
0.5&2&-0.471&0.038&0.216\\
1&0&-0.846&0.083&0.338\\
\hline
\end{tabular}
\end{center}
\label{tquartic}
\end{table*}

The results in Table \ref{tquartic} indicate that the power per length is not determined by the relative size of the quartic term in comparison to the quadratic term. It appears that the power per length is determined by the ``global correlation,'' which we define as $\prod_{j=1}^N [|\langle\ps_j^*\exp(-2\pi i\Phi /N\Phi_0)\ps_{j+1}\rangle |/\langle |\ps_j\ps_{j+1}|\rangle]$. We conclude that, though a large quartic term diminishes the power transfers, it does not eliminate them.

\section{Summary and expectations\label{last}}
Based on the Ginzburg--Landau model for superconductivity, we have evaluated the average local power absorbed (or released) from the fluctuating electric field by the circulating current, in a family of systems that represent the flowing charges in a loop with nonuniform cross section threaded by a magnetic flux. The evaluation was performed following three apparently independent procedures.

Our results are independent of the followed procedure, and violate the requirement that in a system in equilibrium with a heat bath at uniform temperature there should be no heat flow.

The present study analyzes a \textit{gedanken} experiment: we have dealt with a question of principle, and not with technical feasibility. Let us now estimate the performance that the Ginzburg--Landau approach predicts for an aluminium loop of perimeter $L=10^{-6}$m, mean cross section $w_0=10^{-14}$m$^2$ and residual resistivity $10^{-8}\Omega$m. We first check whether the requirements of small coefficients $\alpha$ and $\beta$ are met. The value of $\alpha$ can in principle be chosen by controlling the temperature; the values of $\beta$, according to Table \ref{tquartic}, should not be much larger than $\hbar^2 w_0/mL\sim 6\times 10^{-47}$kg\,m$^5$\,s$^{-2}$, whereas the value of $\beta$ for aluminium is smaller by seven orders of magnitude, indicating that also superconducting materials with stronger coupling would be permissible. The predicted average current is of the order of $ek_BT_c/\hbar\sim 10^{-8}$A, and the power per length is of the order of $10^{-1}(ek_BT_c)^2/\hbar^2\sigma w_0\sim 10^{-10}$Wm$^{-1}$.

Conceivably, the disagreement of the GL approach with the experimental measurements \cite{Crow} for small pair-breaking interaction is related to the thermodynamic failure found here. This disagreement can be amended by addition of the terms found by Maki \cite{Maki} and Thompson \cite{Th}. Thus far we have not investigated whether the addition of correction terms to GL could lead to compliance with thermodynamics.

\section*{Acknowledgements}
The author has benefited from correspondence with Armen Allahverdyan, Denis Basko and Hendrik Bluhm. Part of the computations have used resources of the Technion – Israel Institute of Technology.


\end{document}